\documentclass[a4paper,showkeys,floatfix,aps,pre,reprint,groupedaddress]{revtex4-1}
\usepackage{amsmath,amssymb}
\usepackage{graphics,graphicx}
\usepackage{dcolumn,bm}
\usepackage{psfrag}
\usepackage{color}

\begin{document}


\title{Story of the Developments in Statistical Physics  of  Fracture, Breakdown \& Earthquake: A Personal Account}

\author{Bikas K. Chakrabarti}

\affiliation{Condensed Matter Physics Division, Saha Institute of Nuclear Physics, 1/AF Bidhannagar, Kolkata 700064, India.\\
bikask.chakrabarti@saha.ac.in}


\begin{abstract}
We review  the developments of the statistical physics
      of fracture and earthquake over the last four decades. We
      argue that major progress has been made in this field
      and that the key concepts should now become integral
      part of the (under-) graduate level text books in condensed
      matter physics. For arguing in favor of this, we compare
      the  development (citations) with the same for some other
      related  topics in condensed matter, for which Nobel
      prizes have already been awarded.

\end{abstract}

\maketitle

\section{Introduction} 
When we decided in 1981, working from Kolkata, to investigate
the statistical physics of fracture in disordered solids, our
colleagues in statistical physics could not be very kind to us.
Studies on renormalization group theory of critical phenomena
were at their  peak (Nobel prize to Keneth Wilson next year),
while the friends from the mechanical engineering departments took pity on us as,
though not complete, most were assumed to be reasonably understood  (from
continuum mechanics discussed in standard engineering text
books).

Our motivation had been somewhat ambitious. Asya Skal and Boris
Shklovskii\cite{skal75} in 1975
and Pierre-Gilles De Gennes\cite{degennes76} next year 
had already forwarded their node-link-blob model
of percolation clusters in disordered solids and analyzed the
scaling behaviour of their (classical) linear responses like
electrical conductivity or elasticity. We intended to extend
these studies for non-linear (and irreversible) responses,
like fuse, (dielectric) breakdown or fracture of disordered
solids. We of course realised the major difficulty of the
problem to come from the extreme nature of the breakdown
statics: Unlike the linear responses which get affected by
all the defects (with self averaging statistics), the
breakdown phenomena get nucleated around the ``weakest"
defect (inducing extreme statistics).

Extending the study of the statistical physics of fracture to earthquake statisics had been natural,
though more involved and formidable. Detailed reviews of these developments have been published in review
papers and books, some of whom have been referred to in this account as well in appropriate places. This
account tries to capture those developments since early 1980´s, in which me and my colleagues have been
involved.

\section{Short story of the past: From da Vinci to de Gennes and Mott}

Fracture or breakdown studies might be the oldest
physical science study, which remains still intriguing
and very much alive. Leonardo da Vinci, more than five
hundred years back, observed that the tensile strengths
of nominally identical specimens of iron wire decrease
with increasing length of the wires. (see e.g, ref.\cite{lund00} for
a recent discussion). This is manifestation of the
extreme statistics of failure (bigger sample volume can have
larger defects due to cumulative fluctuations where failures
nucleate and induce lower strength of the sample). Similar
observations were made by Galileo Galilei more than four
hundred years back: ``From what has  already been demonstrated,
you can plainly see the impossibility  of increasing the size
of structures to vast dimensions either in  art or in nature;
likewise the impossibility of building ships, palaces, or
temples of enormous size in such a way that their oars, yards,
beams, iron-bolts, and, in short, all their other parts will
hold together; nor can nature produce trees of extraordinary
size because the branches would break down under their own
weight; so also it would be impossible to build up the
bony structures of men, horses, or other animals so as to
hold together and perform their normal functions if these
animals were  to be increased enormously in height; for
this increase in height  can be accomplished only by
employing a material which is harder and stronger than
usual, or by enlarging the size of the bones, thus
changing their shape until the form and appearance of
the animals suggest a monstrosity"\cite{link}.

For the next 300 years, we did not see major
attention to such problems. In 1921, Alan Arnold Griffith
of the Royal Aircraft Establishment (UK),
estimated how the the crack nucleation stress for an
otherwise pure material decreases with the dimension
of the single defect in the brittle limit (when the
stress-strain relationship remains linear until breaking; inducing the elastic energy density to grow with the square 
of the stress)\cite{griff21}. 
This energy
balance theory for brittle crack nucleation, obtained by
equating the lost elastic energy (proportional to the crack volume) with the surface energy (proportional to the crack surface area)
 of the additional
surface created by further opening up of the defect or
micro-crack, led to a precise estimate of the breaking
strength or stress of the brittle solid, decreasing
with inverse square root of the size or length of
the defect in the direction perpendicular to the stress.
This led to a major development in the study of the
mechanics of brittle fracture.

Subsequently, in 1926, Frederick Thomas Pierce\cite{pie26}
from the British Cotton Industry Research Association in Manchester,
 discovered
what is known today as the Fiber Bundle model, a
fantastically  rich and elegant model to capture the
fracture dynamics in  composite materials. In this model,
a large number of parallel Hooke-springs or fibers are
clamped  between two horizontal platforms; the upper one
helps the bundle hanging while the load hangs from the lower
one. The springs/fibers are assumed to have identical
spring constant though their breaking strengths are assumed
to be different. Once the load per fiber exceeds its
own threshold, it fails and this extra load  is shared
by the surviving fibers. If the platforms are rigid,
there is no local deformation around a failed fiber
(and no stress concentration around the defect created)
and load is shared equally by all the surviving fibers.
Obviously, such a fluctuation-less model allows several
features of its failure dynamics analytically. This was
first indicated\cite{daniel45} by Henry Daniels from the Wool Industries Research Association
in Leeds, in 1945. 
Study of these models led to important
developments, though they were practically confined to
structural engineers for fitting the material failure
data. Physicists did not notice, rather were unaware of,
these models until late eighties or early nineties.

\subsection{Fracture propagation}
Nevill Mott\cite{mott48} of the Cambridge University, in 1948 extended the energy balance method of Griffith to
include the crack propagation energy. This energy (kinetic energy of propagation), along
with the energy of the newly opened up surfaces, should
balance the elastic energy lost due to the crack
propagation. The crack velocity, which had been zero
in Griffith theory, starts growing with the length of
the crack and approaches the sound velocity in the solid
corresponding to elastic modulus of the released elastic
energy. This led to an extensive literature on the
growth of brittle cracks. Particularly, the morphology of the
crack surfaces (out of plane) was claimed to be universal and the
crack dynamics was characterized as a dynamical critical phenomena
(see Ref.\cite{bou97} for an early review). Much was studied later on the 
in plane growth of the crack, starting with the nice experiment from the
Oslo group\cite{maloy} (see also Ref.\cite{epl10}).

\subsection{Extreme statistics \& distributions}
 
It is natural to expect that for randomly disordered solids
 the linear response to stresses or fields, like those given by the elastic moduli,
or the electrical conductivity (of random resistor networks),
will have self-averaging property ensuring that the
(configurationally)  averaged elastic moduli and
conductivity are defined in the thermodynamic limit (unlike in
quantum cases; e.g., the non-self-averaging conductivity due to
Anderson localization). It
is obvious, however, that the same would not be true
for (even classical) nonlinear and irreversible breakdown
properties of disordered solids. The stressed solid sample
would survive (not break or fail) only if all the microscopic
defects (due to disorder) survive under the stress, indicating
that the fracture or breakdown strength of the solid would
be determined by the weakest or extremely vulnerable defect
in it.

As indicated already, the above-mentioned studies
following Griffith-like energy balance concept, had
limitations on several counts. The assumption of
brittleness of the solid, or linearity in stress-strain
relation up to the breaking point, had been one. More
serious had been assumption of a single or dominant
defect in the entire solid volume.  We discussed earlier
(in the context of Griffith law), the strength
of a solid with one isolated defect (or a dominant
defect in an otherwise elastically homogeneous solid, having
non-overlapping stress released regions of the other
microscopic defects)  decreases with the defect size
(inversely with the square-root of the defect length in
the direction perpendicular to the stress, in a brittle
solid).

In presence of random generic defects in a solid, even
brittle one, the stress released regions of the defects
overlap and do not allow a straightforward generalization.
In a randomly disordered solid therefore the probability of
a larger defect due to configurational fluctuation increases
with the volume of the sample. As the survival of the sample
under stress means then survival of the weakest one in the
sample, with increasing volume (with nominally identical
microscopic defect concentration) the fracture or breakdown
strength of the solid sample decreases.

Because of the possibility the existence of bigger
or weaker defects coming from statistical fluctuations
of overlapping neighboring micro-defects, the effective
strength of the solid decreases with increasing volume,
even for nominally identical composition and elastic
behaviour. This cumulative growth of micro-defect
fluctuations, as captured in the ``distribution tail"
argument of Ilya Lifshitz\cite{lif65} induces  extreme statistics of
the failure behaviour of solids: the cumulative failure
probability of such a solid increases to unity as the
stress grows at a fixed volume or as the volume grows,
at any fixed non-vanishing stress.

This non-self-averaging statistics of the breakdown of solids
are well captured in different limits by the extreme statistics
of Waloddi Weibull\cite{weibull51} and of Emil Gumbel\cite{gumbel54} variety.
Microscopic derivations  of these results came much later (see
the next section) and phenomenologically they were fitted
to the celebrated  extreme statistics of  Weibull and Gumbel
(see Ray and Chakrabarti\cite{rc85_2}, published in early 1985, and Chakrabarti and
Benguigui\cite{oup97} for
an approximate microscopic theory, using percolation statistics, to derive
these extreme  statistics of breakdown in solids, employing
the fluctuation model sketched in the earlier para).
Obviously, equating the failure probability to unity, one would 
get from both the distributions,  fracture strength
decreasing with the increasing volume of the sample.  In early winter of
1985, I was visiting Oxford and gave a talk essentially based on the report made in
Ref.\cite{rc85_2}. Phil Duxbury, who just finished his Ph. D., and moving to the US for Post Doc,
was among
the audience. He
immediately realized and asked for a more precise  argument to justify  the
decreasing failure strength with sample volume. Soon, with Paul Leath and
others he
made a beautiful argument using  the Lifshitz tail argument to estimate
the dominant
defect size and the consequent decrease in fracture strength  with volume
of the
sample using the Griffith kind of argument to relate the decreasing
strength of a
solid with increasing defect size\cite{dux86}.

\section{Statistical physics of fracture: Mandelbrot \& others}
\subsection{Fracture surface roughness}

As discussed in section 2.1, Mott initiated the study of
fracture propagation in solids and studied the propagation
velocity (terminal value) in brittle solids. Such calculations
assume that the excess of the released elastic energy over the
crack surface energy (taking flat surface structure) goes to
the velocity dependent kinetic energy of the crack-tip. However,
the roughness of the crack surfaces were too prominent to
neglect, and there were even conjectures that crack propagation
is more like a turbulent motion (rather than streamline) and
the fracture surface roughness captures this frozen turbulence
in crack propagation.

Benoit Mandelbrot and colleagues first analyzed\cite{mandelbrot84} in 1984
the  observed roughness of different fractured surfaces and
suggested a scale-free fractal behaviour. They
measured the growth of out-of-plane fluctuation of the
fractured surfaces for several steel samples, by defining the
average fluctuation in the surface heights at different
distances of separation on the fracture propagation plane, and
found that on average the fluctuations in heights grow with the
distance of separation  along the plane and follows a power law
(does not follow a scale dependent functional form like
exponential or similar functional form) with an universal value
of the power (exponent). This observation of universality,
together with the later extensive ones, confirmed the existence
of critical behavior and statistics in fracture and breakdown
phenomena. This opened up the investigations of critical
phenomena in fracture and breakdown.

\subsection{Fracture of disordered solids: Percolation models}
When Purusattam Ray joined me in 1984 for his Ph. D. research,
I found him bold enough to take up the challenge of exploring
the origin of extreme statistics of fracture and breakdown in
lattice statistical percolation models of disordered solids.
Though the nature of challenge was not realized immediately,
the prospect of any success in the limited period Ph. D.
research was not clear and looked rather frightening! The
idea was first to extend the percolation scaling theories of
random resistor networks or elastic networks of Skal-Shklovskii\cite{skal75} 
and de Gennes\cite{degennes76} type for linear responses like
conductivity and  elastic moduli to that for electrical
(fuse or dielectric) breakdown and fracture of percolating
networks. The next step of (off-lattice) molecular dynamic
simulation of such elastic networks appeared already a
formidable and distant goal, if at all achievable in any
reasonable time frame with the computing facilities
available that time to us! However, the spirit of Purusattam 
was indominable and that encouraged us a lot.

As mentioned already, observation of Mandelbrot et al.\cite{mandelbrot84}
encouraged the view supporting the existence of critical
phenomena in breakdown dynamics. We therefore proceeded
with the node-link-blob model of of the incipient critical
percolation cluster proposed by Skal-Shklovski and de
Gennes (see e.g., Ref.\cite{stauffer94}) to estimate the  scaling behavior of the
fracture stress, as the percolation threshold is
approached, of a fixed sized sample (large but finite,
to avoid the  failure at vanishing stress, due to the
presence of the extremely week defects in the sample).
Here, one could assume that the vulnerable defect size would
be given by  the percolation correlation  length, while the
elastic modulus would have the power law behavior already
established in the node-link-blob model\cite{stauffer94}. One could
also utilize the fractal dimension of the percolating
backbone to find the scaling behavior of the surface energy
density for calculating the fracture stress in the Griffith
model\cite{rc85_2,oup97,rc85_1}. Indeed, Purusattam achieved already the
molecular dynamic simulation of Lennard-Jones systems of
randomly dilute solid initially on square lattice and with
interaction cut-off beyond a distance of 1.6 lattice constant
and up to a modest system size of 400 atoms\cite{rc85_1}. Though
the general trend of decreasing fracture strength with
increasing concentration of initial lattice (site) dilution
could be seen, results for bigger system sizes were needed
for any reasonable analysis.

The paper however attracted attention of several important
groups. Dietrich Stauffer, in particular, invited us to extend
this molecular dynamic study of fracture in disordered lattices
near percolation threshold. Redefining on triangular lattices
(to avoid the shear instabilities) and parallelizing the
simulation program, we were allowed to utilize for more than
seven/eight months the Vector computing facilities in Germany
available to him  that time (using remote log-in and job
submissions etc through telephone from his office in
Cologne!). The results of this study\cite{chakrabarti86}, for system size up to 4225
atoms, clearly demonstrated that, at fixed system size,
the fracture stress monotonically decreases with
increasing dilution concentration and tends to vanish at the
percolation threshold. Also, at any fixed dilution
concentration,  the fracture stress decreases with
increasing system size (as the consideration of extreme
statistics would suggest). This confirmation was  very
intriguing and led to important investigations later. It was
clear, however, the off-lattice molecular dynamic simulations
for disordered elastic networks, undergoing large local
deformations for the nucleation and propagation of fracture
would soon become formidable as the system size is increased
further to check the scaling behaviors.

Hans Herrmann from Cologne and his collaborators immediately
introduced\cite{dearc85} the random fuse
networks, where the local failures or fuses of any lattice
bond would induce modifications in the current distributions
to keep the total current through the network conserved. This
would induce further fuse at the hot spots and the breakdown
would proceed. Since the lattice remains intact (no off-lattice
simulations were required), the computations became much simpler
and universalities in breakdown phenomena could be immediately
checked. When we were struggling so hard  with the molecular
dynamic simulations to extract the universal features of
the breakdown, the fuse model\cite{dearc85} proposed by de Arcangelis
et al. clearly indicated a much softer way to proceed. The
paper came to Stauffer for refereeing, and he made important
comments (including some on the earlier studies) on the manuscript,
which were accommodated in the published version. The model
became an instant success in this field of  investigating
critical behavior of breakdown. It was like the success of
the lattice-gas model over that of the extensive analytical
and numerical (including molecular dynamic simulation)
studies in the  1940-60s to establish the Ising universality
class of the liquid-gas transition at the critical point. We
were indeed awestruck, though chose to continue  our molecular
dynamic studies of fracture in randomly disordered solids for
some more time!  Later, my student Subhrangshu Sekhar Manna
studied the statistical difference, if any, between the minimum
gap (minimum number of dielectric bonds on any path connecting the ends of the sample) and the breakdown voltage
(number of broken dielectric bonds on the breakdown path) in the case of
dielectric breakdown  in the lattice model of random conductor
insulator mixtures\cite{manna87}.  Among others,
this study also triggered several brilliant experimental
investigations on the breakdown behavior  of random resistor
networks. In particular, Lucien Gilles Benguigui of Technion
performed a series of experiments  by employing light-emitting
diodes for insulators in random conducting networks under
large voltage gradient. The failure path could be made visible
by the lighted  diodes (e.g. Ref.\cite{ben88,ben94}, see also\cite{oup97}). 

It is worth noting, however, that Purusattam and coauthor\cite{ray_date} showed that
percolation-like mode of breaking (rather than nucleation-like breaking)
dominates as one increases disorder. Recently Shekhawat et al.\cite{shk10}
claimed from their renormalization group study that the avalanche behavior
seen in the fuse model is unstable for finite disorder and flows to
nucleating failure in large system size limit. A percolation-like
failure mode can be seen for very high disorder limit (Moreira et. al.\cite{herr12}).

Anyway, going back to late 1980's, on invitation from one of the editors, I wrote a
mini-review\cite{chakrabarti88} on these developments on
fracture and breakdown in disordered solids. The journal itself
broke down  and quickly disappeared! However, when David
Bergman (Tel Aviv) and David Stroud (Ohio)  wrote their
review on Physical Properties of Macroscopically Inhomogeneous
Media in volume 46 of the Solid State Physics (Academic Press,
1992), they noted (in pp. 264-267) my mini-review as an ``authoritative" one and
suggested for a detailed one in the same series.  I came
to know of it much later, and then planned immediately and
wrote together with Benguigui, the book Statistical Physics of Fracture and Breakdown inDisordered Systems\cite{oup97}, which was
published from Oxford University Press in 1997.  Muhammad
Sahimi (Southern California) developed further these scaling
studies for disordered solids in a series of papers during
this period and reviewed all these results in a major
compendium\cite{sahimi03} in 2003. Somehow, the choice of
timing of the both these books were somewhat wrong. The major
developments  in the  statistical  physics of fracture in
Fiber Bundle Models started getting settled a little later!

\subsection{Fiber bundle model \& its Statistics}
As mentioned in the Introduction (section 1), the fiber
bundle model was introduced by Pierce\cite{pie26} in 1926 as a
model to understand the strength of composite materials.
The model is deceptively simple: the bundle consists of
a macroscopically large number of parallel hook springs
of identical length and, for simplicity, each having
identical spring constants. They have however different
breaking stresses. All these springs hang, say, from a
rigid horizontal platform. The load hangs from a lower
horizontal platform, connected to the lower ends of the
springs. This lower platform can be assumed to be
absolutely rigid, when the load at any point of time
is shared equally, irrespective of how many fibers
or springs have broken and where, by all the surviving
fibers (equal load sharing model). The lower platform
can also be assumed to have finite rigidity, so that
local deformation the platform occurs wherever springs
fail and the neighboring surviving fibers have to
share larger fraction of that transferred from the
failed fiber. Extreme case is that of local load
sharing model, where load of the failed spring or fiber
is shared (usually equally) by the surviving nearest
neighbor fibers. As may be guessed, the failure
dynamics of the equal load  sharing model is easier to
formulate and analyze. In fact, the strength of such a solid was first estimated
by Daniels\cite{daniel45} in 1945.

In spite the elegance of the model and many profound
features, the model did not catch the attention of
physicists until late eighties in the last century, when
Didier Sornette noted some other attractive features of
the equal load sharing fiber bundle model\cite{sor89}. Later, when
Purusattam explained to us in early 1998 about their
intriguing mean-field study in Gene Stanley's group in
Boston  on the possible first order transition behavior
of  fractures in fiber bundle models\cite{zapperi97}, we were taken by surprise!

Starting a little earlier, when Srutrashi Pradhan joined
me for his Ph. D. research we started  to explore some
simple yet non-trivial versions of the equal load sharing
model. Though these versions were not  of much practical
interest, say, to the engineers, they were expected to
allow us  making more precise  formulation and analysis
of some universal features of its breaking dynamics. The
simplest such a fiber bundle model assumed that the
strength of the fibers in the bundle are uniformly
distributed, starting from zero to a normalized maximum.
It was then easy to set up a simple recursive equation
for the breaking dynamics: when the bundle is loaded
with an external load, all the fibers having strength
up to the value of the load per fiber break and the surviving
fraction would be given simply by the difference of this
load per  fiber from the strength of the strongest fiber
(normalized to unity). However, due to the breaking of
these fibers, the load per surviving fibers increase
exactly by the inverse of the fiber fraction broken in
the earlier step. this increased load per fiber will
induce failure of a further fraction of bonds, and
the surviving fraction of fibers at this stage will again
be given by the  difference of this (increased) load
fraction per fiber from unity (normalized highest
strength). This gives a simple non-linear recursion
relation for the surviving fraction of fibers at any
stage or time (as the load per fiber at
any time is given by the inverse fraction of the surviving
fiber fraction of the earlier step or time).  If there is
a fixed point of the relation at any non-zero fraction of
fibers, then the bundle does not fail under that load
(initially hanged from the lower platform of the bundle),
and the runaway dynamics otherwise  indicates failure
of the bundle. The model was straightforward and the
calculations (even the naturally emerging critical
behavior of its dynamics) was so simple that we first
thought, this must be known already! Srutarshi made an
extensive search and could not find. Just around that
time, we received the acceptance of one of our paper on
the numerical  studies on  precursors of criticality in
some Self-Organized-Criticality models in Physical Review
E. We then made an odd request to the editor
to allow us accommodating a brief section giving
some calculations in a Fiber Bundle model, where such
precursors can also be seen analytically, and also add
that in its title! Surprisingly, the editor readily agreed
and we got the first publication of this model and its
charmingly simple recursion relation capturing the breaking
dynamics in the model\cite{pradhan01}. My student
Pratip Bhattacharyya noted several intriguing features
in the structure of the recursion relations in the model
and a series of studies were made in the following years
(see e.g., Ref.\cite{bhatta03}). 

It was clearly demonstrated in a series of papers (starting with Ref.\cite{pradhan01},
 see e.g., Ref.\cite{pradhan10} for a review) 
that although there occurs a discontinuous jump in
the value of the surviving fiber fraction across the
critical load, they do not signify any first order transition.
This is because,  the failure time, breakdown susceptibility
(given by the ratio of the fraction of failed fibers and
marginal increase in the external load),  etc diverges at
the critical load on the bundle (with mean filed like exponent
values;  due to suppression  of load fluctuations among the
fibers in this equal load  sharing model). The scaling forms
of the relaxation time were later extensively studied in Ref.\cite{roy13}

Unlike in the brittle fractures, where essentially a single
(weakest) crack chooses to nucleate and propagate throughout
the sample (as in the fiber bundle model with local load
sharing), incremental failures throughout the sample, giving
rise avalanches, occur in such equal load sharing fiber bundle
models. With uniform distribution of the fiber strengths, as
discussed above, the power law exponent value for the size
distribution of the avalanches was already argued precisely by
Per Hemmer and Alex Hansen from Trondheim in their classic paper\cite{hemmer92} 
in 1992. This universal value of the avalanche size
distribution clearly fitted the critical nature of the
breakdown statistics in the equal load sharing fiber bundle
model (see e.g., Ref.\cite{pradhan10,hemmer16}).

When Srutarshi joined the Trondheim group for his post-doctoral
work, they together essentially established analytically the structures
of the pre-failure and post-failure dynamics of the equal load
sharing fiber bundle models (mostly discussed in Ref.\cite{pradhan10}). Some
of the intriguing signatures of dynamic precursors in the
statistics of an over-loaded fiber bundle were discovered
later (see e.g., Ref.\cite{hemmer16} for discussions on them).

My student Amit Dutta, together with his student in the Indian
Institute of Technology, Kanpur, studied the fiber bundle model
with discontinuities in the threshold distribution\cite{dutta08}.
They found universal critical exponents, except for the avalanche
sizes, which shows non-universal statistics.

The fiber bundle model is a so called toy model. Though it captures
the essential dynamical feature of load sharing following a failures and 
the subsequent dynamics, it lacks many other
realistic features of fracture in solids. It has therefore received
more than its fair share of criticism in its early physics-entry stages mainly
from the referees (who could eventually be overruled in most cases).
 However, it is worth noting that it is the simplicity
of the model that gives rise to immense flexibility and hence could
be applied to  diverse topics such as power-grid networks, failures
in ice blocks, traffic jams and of course fracture of disordered solids. 
Such advantages of flexibility and potential for diverse applications
were exploited in many cases. Particularly, in Ref.\cite{sen_biswas}, 
a limiting strength for the system under uniform loading but non-uniform
load sharing was derived. Although not directly applicable to fracture,
in other systems such as power grids, non-uniform load sharing could
be interesting.

The main difference between the fiber bundle model and other models of 
fracture (e.g. fuse model) is that in fiber bundle model, the range of
load redistribution is also a parameter at hand. While the critical behavior 
in the equal load sharing
model (mean field limit) was mostly studied, the local load sharing rule\cite{harlow78,rbr} gives
nucleation driven extreme statistics (a crossover occurs near the percolation
threshold, when these two rules are mixed; see discussions in section 4.2 following the Ref. \cite{biswas13a}).  In Ref.\cite{rbr} a
full phase diagram in range of redistribution and strength of disorder 
was estimated and presented. This shows the various modes of failures observed in the
model over the years in different parts of the phase diagram.

\section{Statistical physics of earthquakes: Omori \& Gutenberg-Richter}
That earthquakes are large scale dynamical breaking
phenomena, occurring due to  the stick-slip kind of
failure at the earth-crust interface with the slowly
moving tectonic plates, had been known for a long time.
Two major power laws in the statistics of earthquakes
had clearly indicated the possibility of criticality
in their dynamics. Long back in 1895, Fusakichi Omori
of the Tokyo University suggested that the rate of
the aftershock counts decreases inversely with time
elapsed since the main shock at any epicentre\cite{omori}.
Utsu later modified that law saying that the rate
of aftershocks decrease inversely with a power of
the time plus an adjustable constant and the power is
close to unity\cite{utsu} and varies in the range 0.7-1.5. 
Beno Gutenberg and Charles Francis Richter, both from Caltech, in 1956 proposed
a law saying that the logarithm of the number of
earthquakes of a particular magnitude or more,
occurring in a given region and time period
decreases linearly with that particular magnitude.
Equating the log of the energy released in an
earthquake linearly with its magnitude (as often
confirmed in underground nuclear blasts of known
energy and the consequent seismic magnitudes),
one gets a power law relation between the
earthquake frequency and the energy released. More
specifically, the number of earthquake events
releasing a particular amount of energy or more,
in any area or period, decreases with an inverse
power of that particular energy\cite{gr}.

\subsection{Burridge-Knopoff model and its statistics}
As mentioned earlier, these scale free form (power
laws with universal values of the exponents) of
the earthquake statistics immediately indicated
the possible role of the underlying critical
phenomena in the dynamics. One of the earliest
and so far the most successful model for
earthquake was proposed\cite{bk67} by Robert Burridge (Univ.
Cambridge) and Leon Knopoff (Univ. California
Los Angeles), in 1967 
(see also Ref.\cite{bur06} for a detailed discussion on the model).
One takes a chain of a large number of wooden blocks
connected by Hook springs placed on a rough horizontal
table. One end of the chain is free and the other end
is pulled horizontally by a motor. The other end of
the chain is kept free. The rough surface contacts
between the wooden blocks and the table top would
mimic different portions of the earth's crust
and the tectonic plates. The plate motion (in a
reverse way) is captured essentially by the motion
of the chain induced by the motor pull.  Though
the motor pull would be uniform, the chain would
have stick-slip type motion; As the static friction
force is higher than while in relative motion
(essential source of non-linearity in the dynamics
of the otherwise harmonic chain), different number
of blocks will slip (different amounts of elastic
energy of the inter-bloc springs will be released)
at different points of time. A motion picture of the
block positions would allow calculation of elastic
energies of the inter-block springs and thereby of
the entire chain or ``train" as the dynamics
progresses from  an initial ``charging" state to
a steady one. The decrease in the number of
bursts with increase in the amount of energy
released in those bursts clearly indicated a power
law, as suggested by the Gutenberg-Richter law.

James Langer (University of California, Santa
Barbara) and collaborators, in a series of papers
published over a decade starting mid-eighties,
formulated a simple version of the Burridge-Knopoff
model using numerical tricks. Here the equation of
motion of each block has a  part of the forces
coming from the relative displacements of the
neighbouring blocks connected by Hook springs,
and a nonlinear part depending on the relative
velocity of the block compared to the table top.
Extensive simulations indeed showed the
Gutenberg-Richter like behavior of the (elastic)
energy burst statistics. A summery of their results
were published\cite{cls94} in a nice review in 1994. 
Hikaru Kawamura of the
Osaka University, and Takahiro Hatano of the Tokyo
University and their  collaborators made extensive
simulation studies on a similar numerical version
of the Burridge-Knopoff model, with more realistic
friction forces etc. It may be mentioned here,
none of these model studies could reproduce the
Omori law for aftershocks.  A review of those
studies  and of other statistical physics
models was published together with us in an
extensive review on the statistical physics of
earthquake dynamics in 2012 in Reviews of Modern
Physics\cite{kawamura12}. 

In the summer of 2012, Soumyajyoti, Purusattam
and myself were attending a fracture meeting in the
SINTEF Petroleum Research, Trondheim, organized
by Srutarshi. One evening, while discussing in the
guest house there, Soumyajyoti and Purusattam came
to a novel computationally simpler Burridge-Knopoff
type model, where the block motions are discretized
(by the lattice structure of the underlying table)
and more importantly, the difficult-to handle
non-linear friction force is replaced by random
(threshold type) pinning forces. Though, no analytic
calculation could be done, Soumyajyoti, on  return
to Kolkata, made extensive numerical studies and
the results showed extremely encouraging features
in the avalanche statistics: both the Gutenberg-Richter
law as well as Omori law were reproduced\cite{brc13} (see
also Ref.\cite{biswas15}).

\subsection{Self-organised criticality: Bak \& others}
The power laws in the distribution functions observed
in nature, like the above mentioned Omori or
Gutenberg-Richter  laws and the universal values of
those powers, clearly indicate the presence of some kind
of self-similarities or scale independent features
in such complex dynamics.  Such  self-similarities
keep the power invariant, like the fractal dimensionality
of the effective space of dynamics. The distribution
function may, if we wish, be viewed as  an effective
``volume" in such a self-similar (fractal) space (geometry)
and it varies with the  event size (viewed as some
effective  inverse ``length"). In any geometry, the power law
relation between  the length and volume follows  naturally.
Changes in the length scale would result in the change in
the volume (in that embedding geometry) by a corresponding
power law, with the power given by the (fractal) dimension
of the space or geometry.  These observations therefore
clearly indicate the role of critical phenomena in earthquake
statistics. However, in the cases of liquid-gas or
ferromagnet-paramagnet phase transitions, where criticality
occurs at specific points,  the systems need to be
brought to the critical point by tuning externally the
(thermo-) dynamic parameters like temperature, etc. Here, in
the example of earthquake we are considering, the system
seems to be self-tuned to criticality!

Per Bak (from Copenhagen) and collaborators proposed in 1987
a toy model, called the sand pile model, which dynamically
evolves towards such a self-organized critical state and
continues its dynamics there without any tuning\cite{btw}. 
 Imagine a horizontally placed square
lattice of finite but large size (having boundaries), where
on any randomly chosen site one  throws unit height (sand
grain). The process of throwing heights on randomly 
chosen sites of the lattice (adding sand to the pile goes
on at a constant but slow rate, much slower than the
dynamics for local failure or toppling discussed next). The
dynamics of (local) failure is such that if the height at
any site becomes four at any time, the site topples (height
becomes zero at that site)  and each four of its neighbors
receives one unit of height. If that causes the height of
any of the neighbor to become equal to four, that site
topples in the next time unit and each of its neighbors
(including the neighbor whose toppling caused its own).
This happens every where, except for sites on boundary,
where the share of the height for the neighbor(s) beyond
the boundary leaves the system (the total mass or height
at any time leaving the system contributes to the size
of the avalanches). Needless to mention that, although the
(input) addition of sand grains or heights to the system
occurs at a constant rate, the (output) rate of mass or
height ejections from the system occurs in bursts or through
avalanches. Numerical studies show that after some initial
``charging" period, when the average height at any site
reaches to about 2.1 (for square lattice and with 4 as the
threshold height at any site, as in the example above), the
dynamics stabilizes to a self-organized critical state
where the avalanche frequencies decrease with its mass or
size following an universal power law with an exponent value
around 1.3 (independent of the lattice or threshold details)\cite{btw}. 
Several extensions of the model were proposed
immediately afterwards to make the model more realistic.
However, they all led to the same universality class
for the critical behavior of their statistics. 

Subhranshu
studied a novel stochastic version of the toppling dynamics
in a computationally efficient version of the model, where
the threshold height becomes 2 and after the site topples,
two neighbors are chosen randomly of the four neighbors,
and they get one unit of hight. If any of these two
chosen neighboring sites had one unit of height earlier,
that site also topples in the next instant, and so on.
I was visiting Forschung Zentrum Julich in the summer of
1990, where Shubhrangshu explained me the model and results.
Because of the stochastic nature of failed load (height)
sharing and the stable values of load or height any point
having binary  values (0 or 1), computationally the model
had been much more efficient and the numerical results seemed
to suggest a new universality class\cite{manna}. The model, now known
as Manna model, has since been extensively studied and a new
(Manna) universality class for such dynamical critical
phenomena is more or less established.
 A more realistic version of the
model for earthquake was proposed by Olami et al.\cite{ofc92}, 
where each instant a toppling
occurs at any site, the entire load (force) is not
shared by the neighbors, but a fraction is assumed to
be lost and dissipated locally. As mentioned earlier
(see section 3.5) early 2012, Soumyajyoti discovered a
brilliant version  of a two  dimensional fiber bundle model
with local load sharing, which was shown to possess
interesting self-organized critical behavior.  In the model,
a horizontally held two dimensional network of hook springs,
having random breaking thresholds, is pulled downwards from
a central site  at a constant rate using a motor. As with
time more and more fibers break, they immediately join
the pulling sting leaving the springs beyond the periphery
of the central defect patch unaffected. All the springs on
the growing periphery of the central broken patch share
equally the constantly growing central pulling force.
This dynamic equilibrium has interesting critical
statistics of failure\cite{biswas13b} (see also Ref.\cite{biswas15}), 
particularly because there was no externally imposed dissipation
scale, but dissipation came from the increase of the effective system size.

Such models therefore provide  natural and generic ones
for explaining the Gutenberg-Richter type universal
behavior of the released energy bursts or avalanches in
earthquakes.  One may note, as such, these models can not
distinguish between the main shock and aftershocks and
therefore they do not capture at all any Omori type behavior
of the aftershocks.


\subsection{Two fractal overlap model}
As discussed in section 3.1, the fracture surfaces have
well-established self-similar geometries and resultant
scaling properties. As earthquakes occur due to the
slips of the rough crust surface over the moving tectonic
plate surface, one can model the earthquake time series
by counting the changes in the measure of the overlaps
of two fractal surfaces, as one of them moves with a
fixed velocity over the other. One can assume that the
elastic energy stored during sticking period in the
 interfacial contacts between the crust and creeping
tectonic plate (measured by the overlaps) gets released
as slip occurs. The time series of these energy bursts are
then given by the time series of these overlaps between
two fractals, as one moves with constant velocity over
the the other.  This maps the entire earthquake dynamics
into a geometric model of finding the two-fractal (mass)
overlap time series. Analytical results for the simplest two
Cantor set overlap series showed not only the Gutenberg-
Richter type law, but also a built-in Omori law\cite{tfo} (see
also Ref.\cite{kawamura12} and Ref.\cite{biswas15} for details). 
Indeed Srutarshi started his research career with study of ``two
fractal overlap model", as part of his Post MSc project and later published a
detailed numerical study on it with Purusattam and others\cite{phy_scrpt}.

\section{Comparison of activities with Those in Other Contemporary topics of condensed matter physics}
\begin{figure}[h]
\centering
\includegraphics[height=4.0cm]{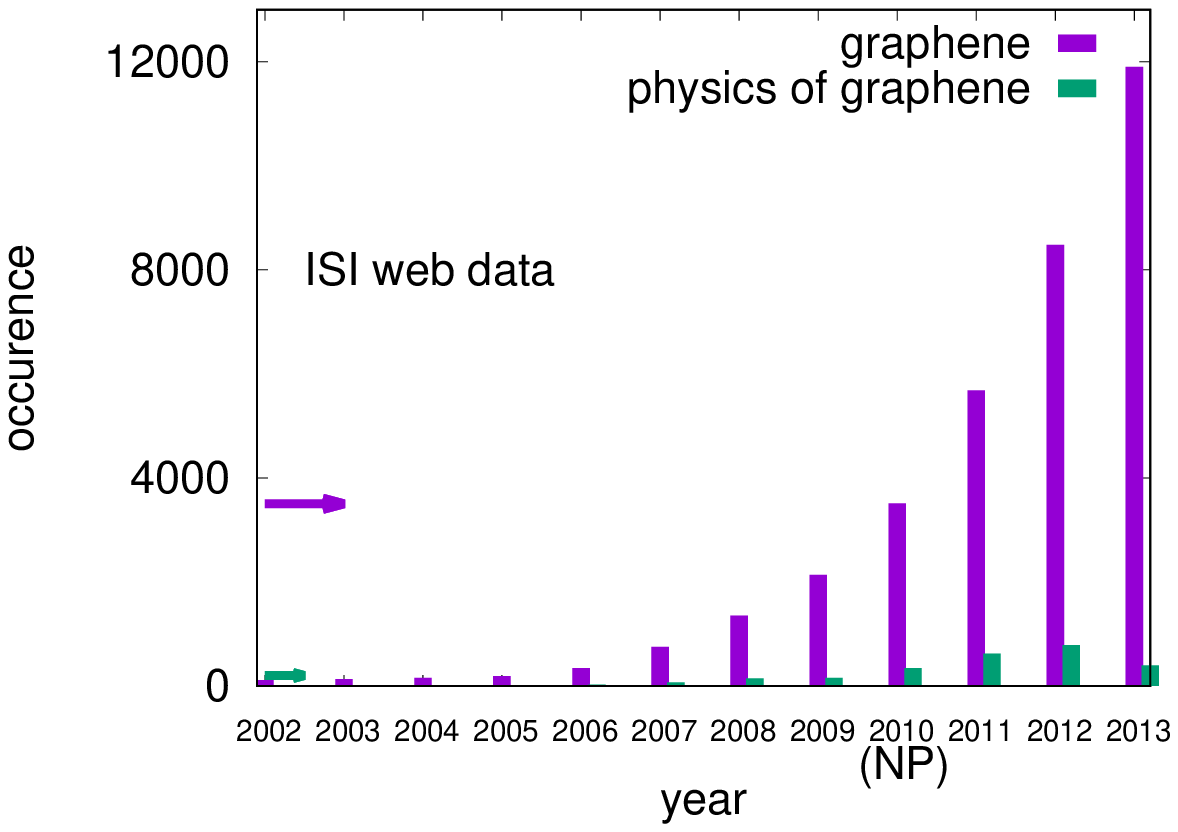}
\includegraphics[height=4.0cm]{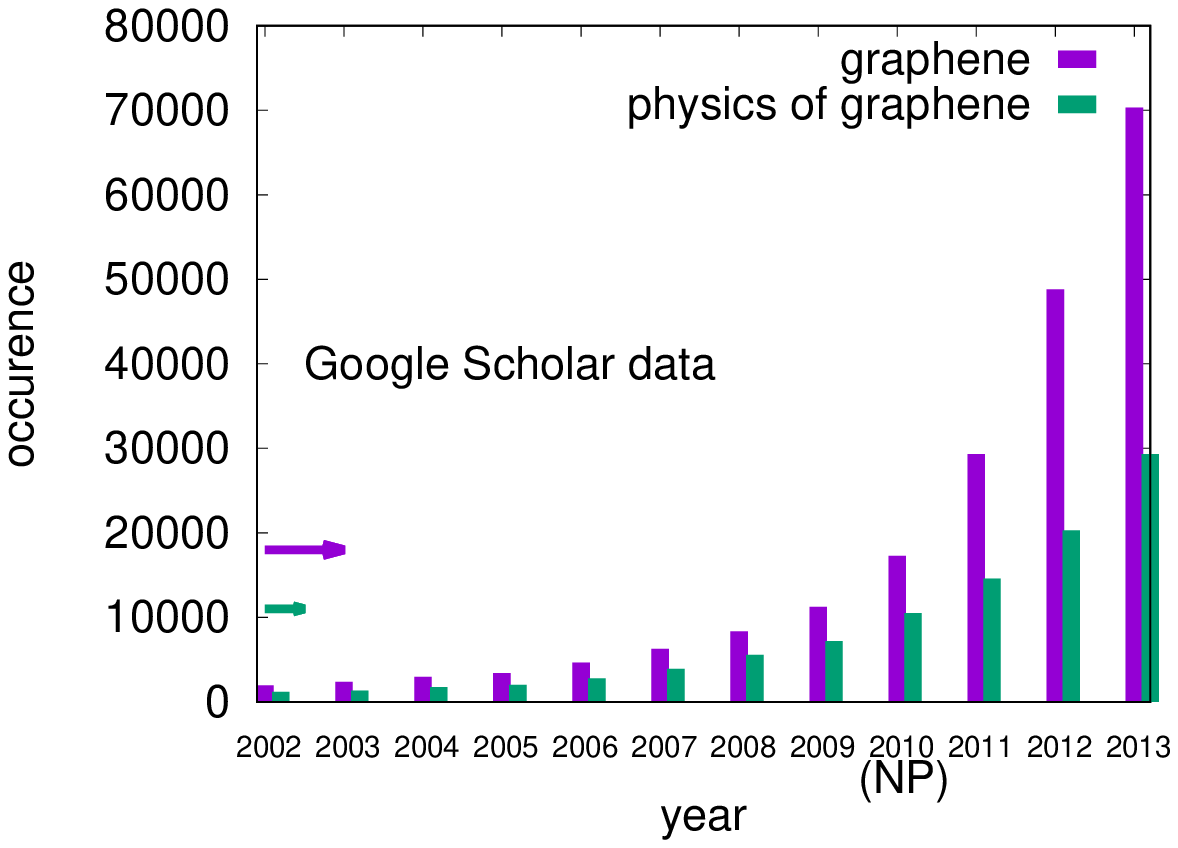}
\caption{(Left) The number of papers on graphene listed in ISI Web of Science around the year the Nobel prize was awarded 
in graphene (2010); on the right the same is shown for the Google Scholar data.}
\label{gr_isi}
\end{figure}

\begin{figure}[h]
\centering
\includegraphics[height=4.0cm]{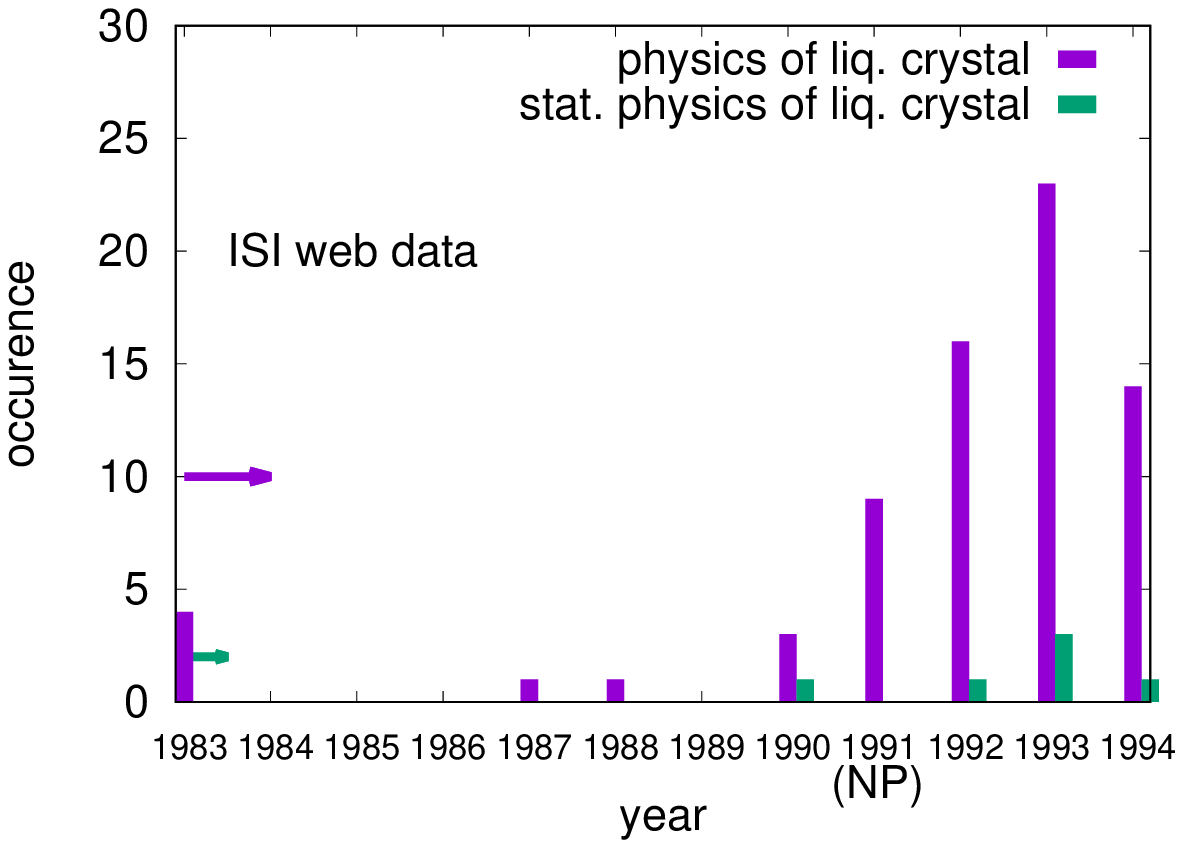}
\includegraphics[height=4.0cm]{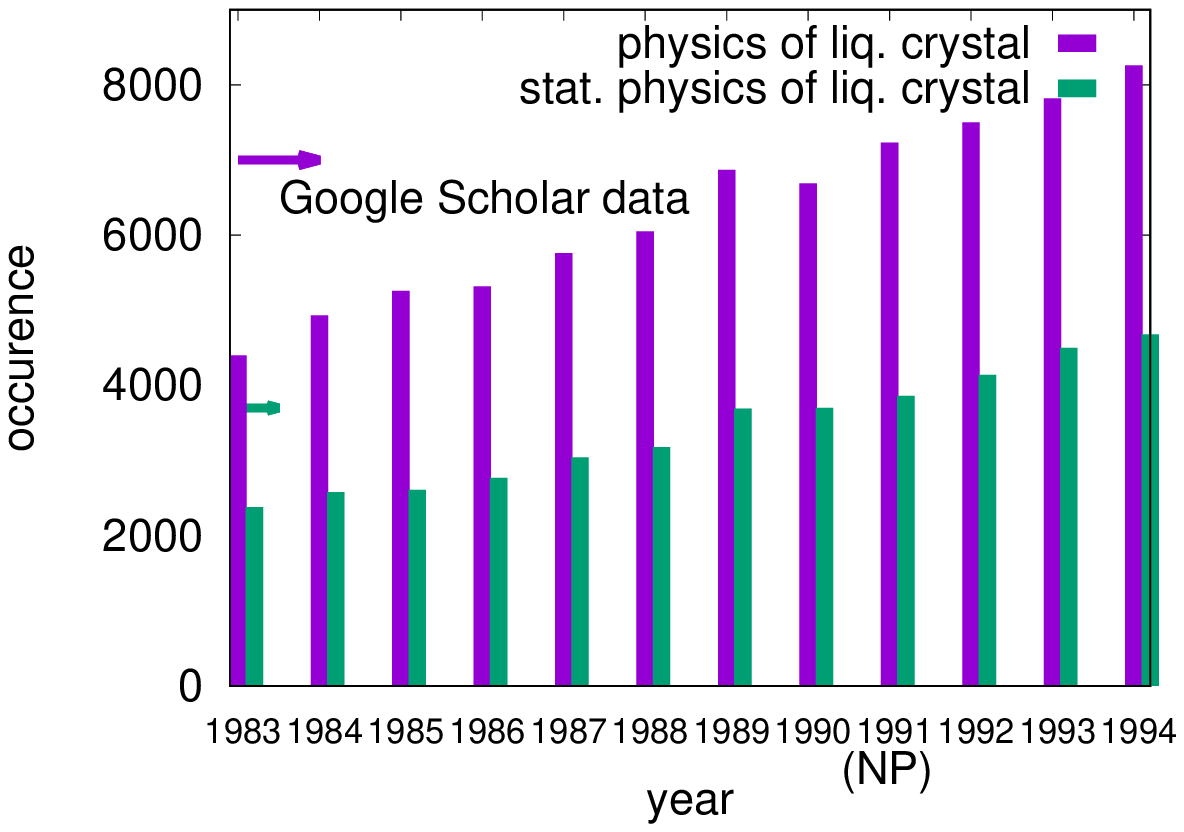}
\caption{ (Left) The number of papers on liquid crystal listed in ISI Web of Science around the year the Nobel prize was awarded 
in liquid crystal (1991); on the right the same is shown for the Google Scholar data.}
\label{lq_isi}
\end{figure}

\begin{figure}[h]
\centering
\includegraphics[height=4.0cm]{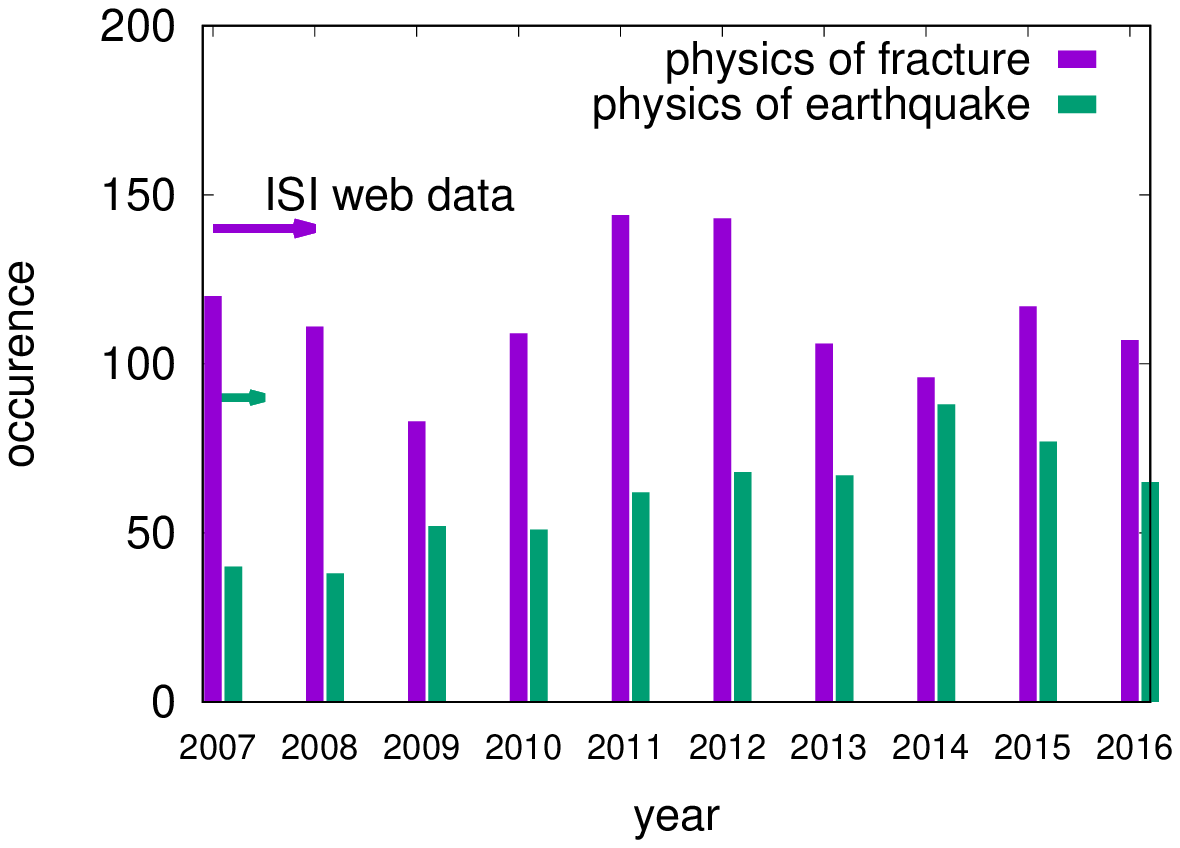}
\includegraphics[height=4.0cm]{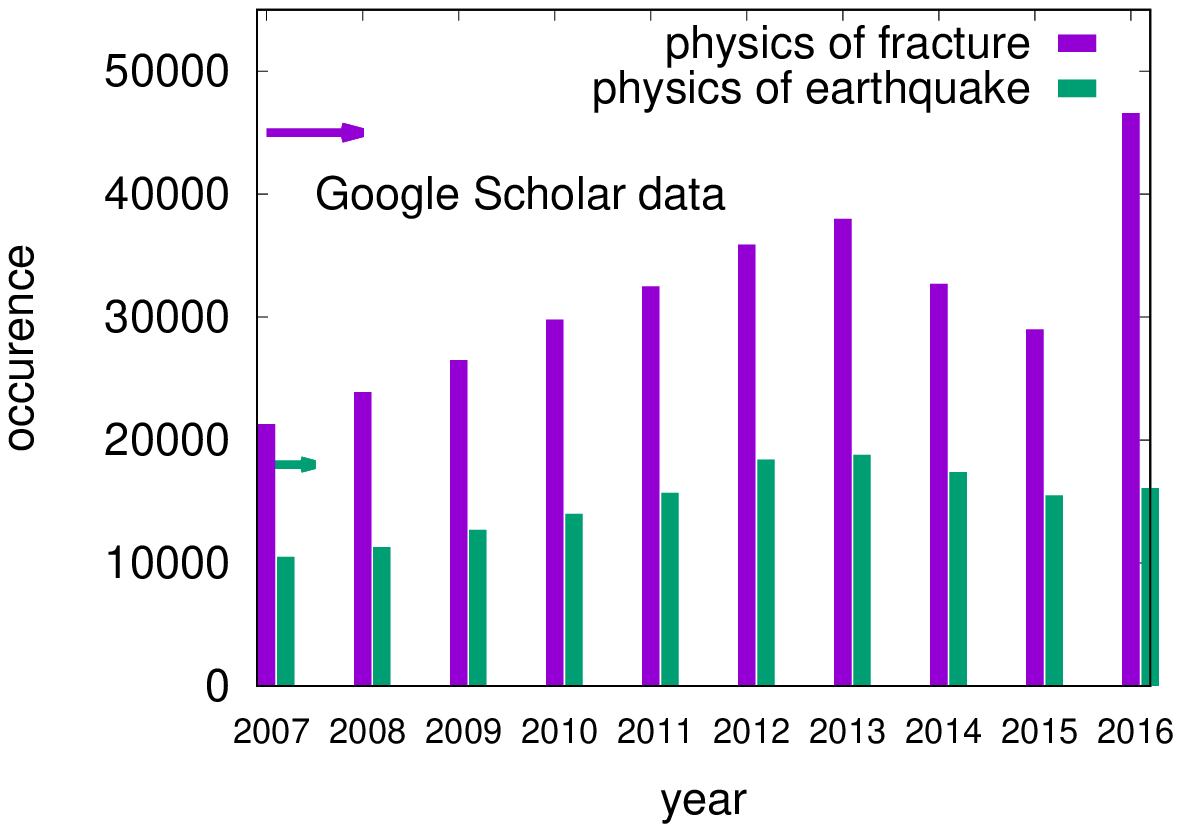}
\caption{(Left) The number of papers on physics of fracture and earthquakes listed in ISI Web of Science;
on the right the same is shown for the Google Scholar data.}
\label{fr_isi}
\end{figure}

\begin{figure}[h]
\centering
\includegraphics[height=4.0cm]{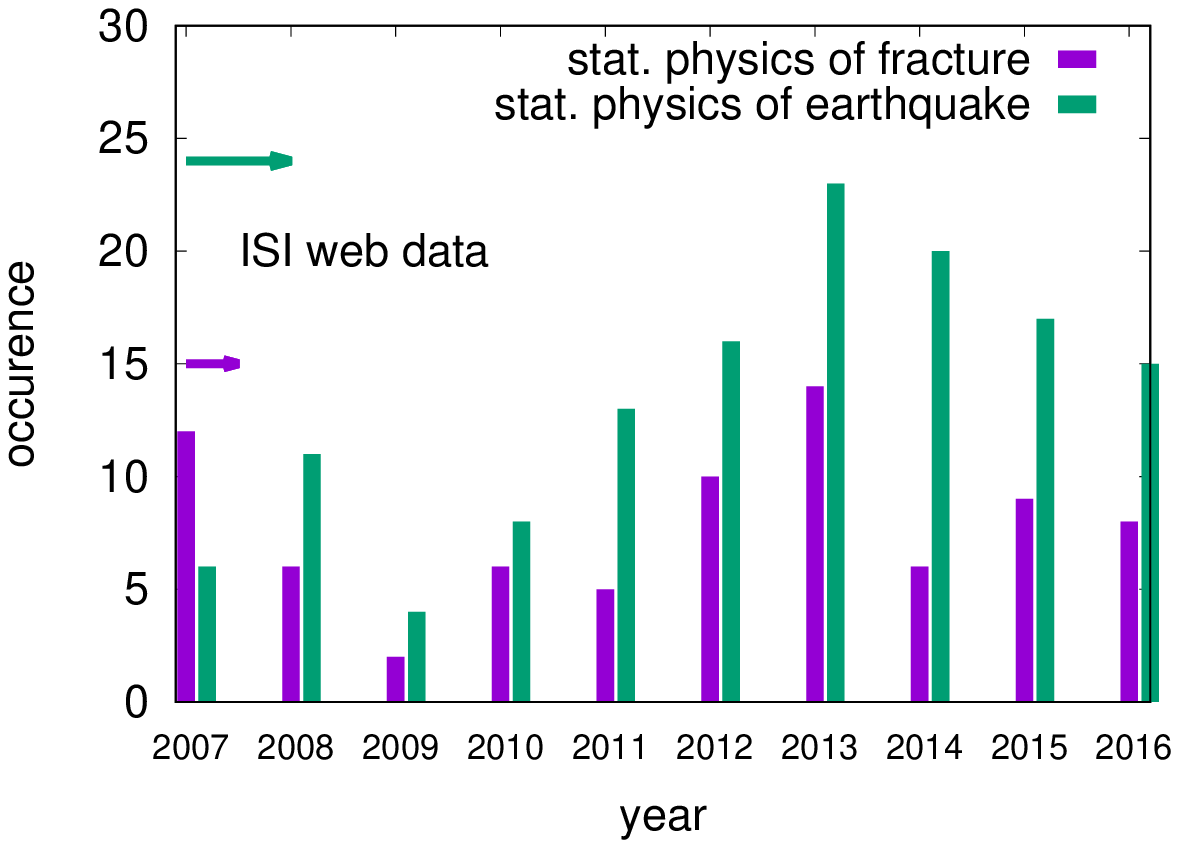}
\includegraphics[height=4.0cm]{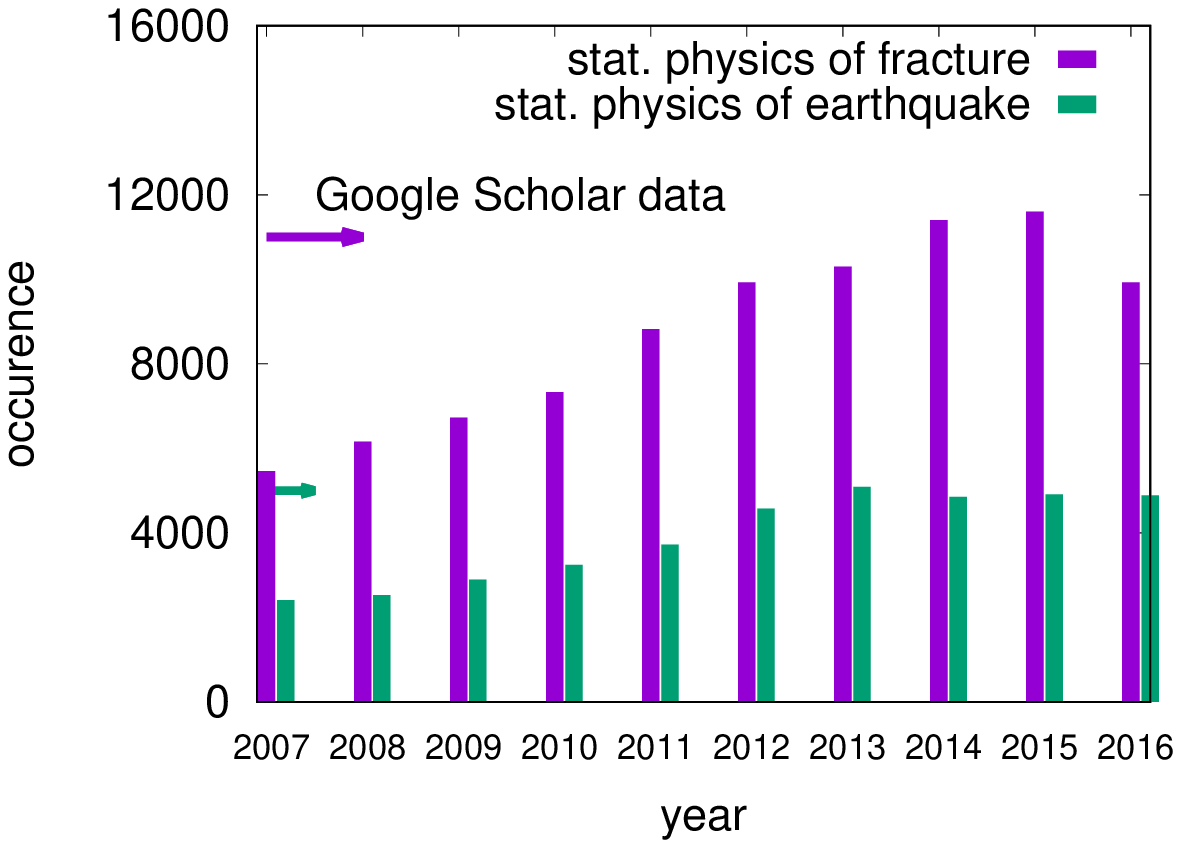}
\caption{(Left) The number of papers on statistical physics of fracture and earthquakes listed in ISI Web of Science; on the right the same is shown for the
Google Scholar data.}
\label{st_isi}
\end{figure}

In this section we wish to compare how the activities in the (statistical) physics of fracture and earthquake compare 
with other branches of condensed matter physics that are considered generally mature (with being awarded Nobel prizes). 
One objective way to compare is to look at the number of papers mentioning the subject unambiguously in the topic
of a published paper (ISI Web of Science data). We compare the data from websites such as Google Scholar, ISI Web of Science (data compiled in January 2017; 
 Figs. \ref{gr_isi}, \ref{lq_isi}, \ref{fr_isi}, \ref{st_isi}). 

Figs. 1 and 2 show the number of papers published each
year on the topics graphene and liquid crystals,
respectively, around the years in which the Nobel prizes
were awarded. Both ISI Web of science data (published
papers; topics search) and Google Scholar data (term
anywhere in published/unpublished documents in the
internet). Graphene data are much higher than those for
physics of graphene data.  Statistical physics of
graphene is not an appropriate topic for search (unlike
for all the rest of the data sets shown in Figs 2-4).
Still, the data are shown, just to indicate the scale
of research activities in the field at the time
of recognition for such a popular and contemporary
condensed matter physics topic. The physics and
statistical physics of liquid crystals are of course
much more appropriate to compare with the corresponding
data (from both ISI Web of Science and Google Scholar)
for fracture as well as earthquake, and are shown in Figs. 3
and 4.

As might be noted, the contemporary rates of publications
(research activities) in both  physics and statistical
physics of fracture and earthquake are quite comparable
and even higher than the respective rates for liquid
crystals research activities around the year of its
recognition. The data for contemporary research activities
in physics of fracture and of earthquake are also
comparable to those for physics of graphene around the
year of its recognition.

\section{Perspectives \& concluding remarks}
The data shown in Figs. 1 through 4 clearly indicate that the progress in the
               studies on the statistical physics of fracture and
               earthquake and their impact in the contemporary
               literature has already been extremely significant.
               It is therefore unfortunate that the standard
               condensed matter physics graduate courses and
               researches do not include even minimal
               discussions on physics of fracture (introduce, say,
               the elegant and versatile Fiber Bundle Model, used
               and explored extensively by engineers and physicists)
               and of earthquake (introduce, say, the Burridge-
               Knopoff model). It may be mentioned that four recently 
published books, namely on ``Earthquakes: Models, Statistics, Testable Forecasts"\cite{kagan_book},
``Desiccation Cracks \& their Patterns"\cite{lucas_book}, ``Fiber Bundle Model: Modeling Failure in Materials"\cite{hemmer16},
and ``Statistical Physics of Fracture, Breakdown \& Earthquakes"\cite{biswas15}, in the series ``Statistical Physics of
Fracture and Breakdown" by Wiley-VCH and edited by Purusattam and me tried to capture all these developments in details. A
suitably picked and chosen set of topics from these set of books can be utilized to generate an appropriate graduate level course. 
We do believe, such a course would be very timely and has been long overdue.

\section*{Acknowledgement}
This is an expanded version of the presentation in the 5th event (January, 2017) in the series of the international conferences on statistical physics of fracture (Fracmeet)
held in the Institute of Mathematical Sciences, Chennai organized by Purusattam Ray and his colleagues. Several encouraging suggestions and comments from the participants
helped me to complete this story. I am thankful to Soumyajyoti Biswas of the Max Planck Institute for Dynamics and Self-Organization, Goettingen and Editorial board member
of this journal to encourage me to finish the account and publish. I am also grateful to the Editor, Ophir Flomenbom, 
for a careful reading of the entire manuscript, thoughtful suggestions, appreciations and ready support. I am grateful
to the J. C. Bose National Fellowship (DST, Govt. India)
 grant for supporting  the publication of this paper. 

\end{document}